\documentclass[epjCONF]{svjour}
\usepackage{graphicx}
\usepackage[varg]{txfonts} 
\usepackage[latin1]{inputenc}
\session-title{MESON2012 - 12th International Workshop on Meson Production, 
Properties and Interaction}
\begin{document}
\title{SIDDHARTA impact on $\bar{K}N$ amplitudes\\used in in--medium applications}
\author{A.~Ciepl\'{y}\inst{1}\fnmsep\thanks{\email{cieply@ujf.cas.cz}} \and J.~Smejkal\inst{2}}
\institute{Nuclear Physics Institute, 250 68 \v{R}e\v{z}, Czech Republic \and 
Institute of Experimental and Applied Physics, Czech Technical University in Prague, 
Horsk\'{a} 3a/22,\\ 128~00~Praha~2, Czech Republic}
\abstract{
We have performed new fits of our chirally motivated coupled--channels model for meson--baryon 
interactions and discussed the impact of the SIDDHARTA measurement on the $\bar{K}N$ 
amplitudes in the free space and in nuclear medium. The kaon--nucleon amplitudes generated by 
the model are fully consistent with our earlier studies that used the older kaonic hydrogen 
data by the DEAR collaboration. The subthreshold energy dependence of the in--medium 
$\bar{K}N$ amplitudes plays a crucial role in $\bar{K}$--nuclear applications.
} 
\maketitle

The recent measurement of kaonic hydrogen characteristics (the $1s$ level energy 
shift and width due to strong interaction) by the SIDDHARTA collaboration \cite{2011SIDD} 
tightens the constraints put on the theoretical models used to describe the low energy 
$\bar{K}N$ interactions. Unlike the older DEAR data \cite{2005DEAR} the new SIDDHARTA 
measurement is also consistent with the $K^{-}p$ scattering length determined 
from the scattering data \cite{1981Mar}. In this short report we examine the impact of the SIDDHARTA 
measurement on the $K^{-}p$ and $K^{-}n$ elastic amplitudes in the free space 
and in nuclear medium. The energy dependence of the $K^{-}p$ amplitudes was 
already discussed in \cite{2012CS} and here we extend the analysis by showing 
our results for the $K^{-}n$ elastic scattering amplitudes as well.

The modern treatment of kaon interactions with nucleons at low energies is based 
on effective field theory, the chiral perturbation theory ($\chi$PT), combined with coupled 
channels techniques used to deal with divergencies that thwart the convergence 
of the $\chi$PT expansion. We employ a chirally motivated separable potential 
model \cite{2010CS}, \cite{2012CS} that matches the effective meson--baryon 
potentials to the chiral meson--baryon 
amplitudes obtained up to the second order in the $\chi$PT expansion in meson 
momenta and quark masses. The model parameters (chiral Lagrangian couplings, 
the low energy constants, and the inverse ranges that define the off-shell 
form factors) are standardly fitted to the $K^{-}p$ low energy cross sections, 
to the $K^{-}p$ threshold branching ratios and to the $1s$ level characteristics 
of kaonic hydrogen. Interestingly, the lowest order (LO) Tomozawa--Weinberg interaction 
alone is quite sufficient to provide a reasonable fit to the experimental data, 
though the next--to--leading order (NLO) is relevant to achieve their good reproduction. 
An important feature of the theory is an energy dependence of the chiral SU(3) 
couplings that bind the various meson--baryon channels. Specifically, the coupling 
of the $\pi\Sigma$ and $\bar{K}N$ channels plays a major role for the dynamics 
of both meson--baryon states. It leads to an appearance of two dynamically 
generated isoscalar resonances \cite{2003JOORM} that are assigned to the $\Lambda(1405)$ resonance 
observed in the $\pi\Sigma$ mass spectrum just below the $\bar{K}N$ threshold. 

In Ref.~\cite{2012CS} we discussed results obtained with three different models 
and showed that all of them reproduce well the available experimental data for 
low energy $K^{-}p$ interactions and the $\pi\Sigma$ mass distribution 
in the $\Lambda(1405)$ region. The models are (see \cite{2012CS} for details):
\begin{itemize}
\item[TW1] - only Tomozawa--Weinberg interaction considered, just two parameters 
(inverse range and meson decay constant, both of them common to all channels) 
fitted to the experimental data including the SIDDHARTA measurement
\item[NLO30] - all LO and NLO interaction terms considered, meson decay constants 
fixed on their physical values, three inverse ranges and four NLO $d$--couplings 
fitted to the experimental data including the SIDDHARTA measurement
\item[CS30] - an older LO+NLO model \cite{2010CS} whose parameters were fitted 
to the DEAR data instead of the new SIDDHARTA ones
\end{itemize} 
The inclusion of the CS30 model allows us to compare the results obtained 
in the DEAR era with those refined due to the SIDDHARTA data. However, 
it was noted in Ref.~\cite{2012CS} that the kaonic hydrogen characteristics 
computed with the CS30 model are already in a better agreement with the SIDDHARTA 
data rather than with the DEAR ones that were used in the fit. This fact 
underlines the much better consistency of SIDDHARTA with the other low energy 
data on $K^{-}p$ scattering and reactions.  

\begin{figure}
\begin{center}
\includegraphics[width=0.78\textwidth]{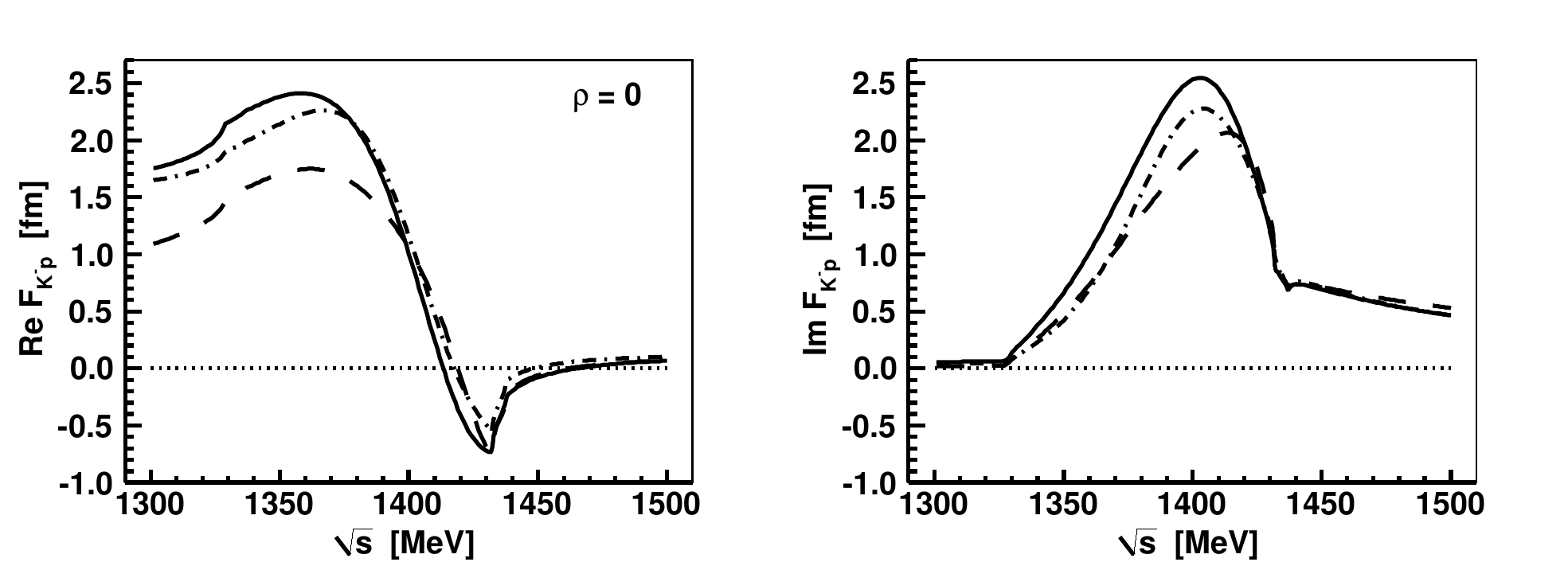}\\
\includegraphics[width=0.78\textwidth]{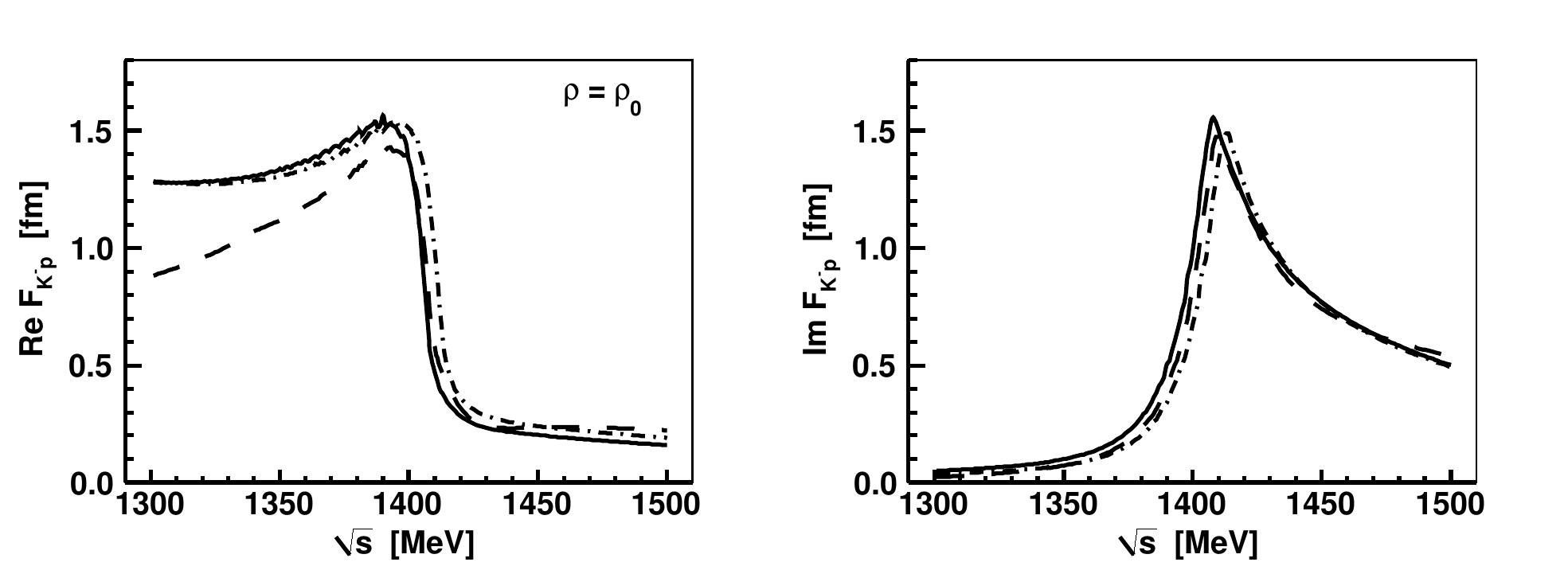}
\end{center}
\caption{Energy dependence of the real (left panels) and imaginary (right panels) parts of 
the elastic $K^-p$ amplitude in the free space (top panels) and in the nuclear medium 
(bottom panels). Dashed curves: TW1 model, dot--dashed curves: CS30 model, 
solid curves: NLO30 model.}
\label{fig:Kp}      
\end{figure}

\begin{figure}
\begin{center}
\includegraphics[width=0.78\textwidth]{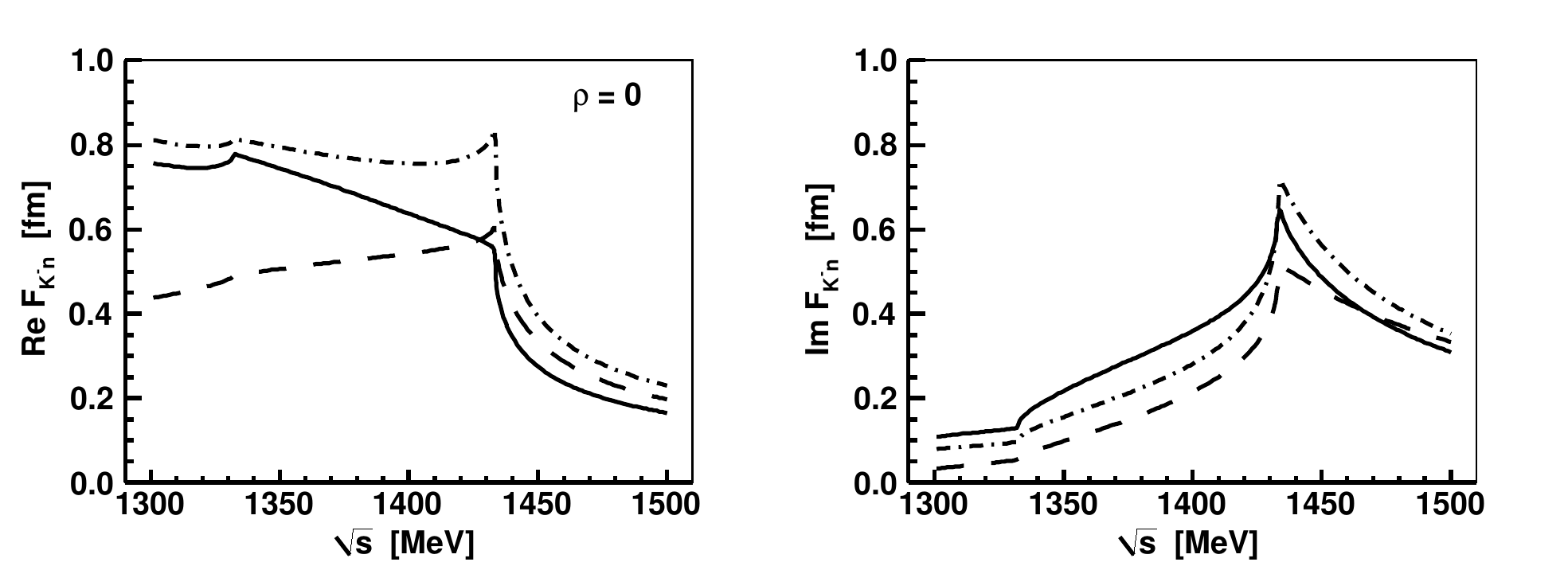}\\
\includegraphics[width=0.78\textwidth]{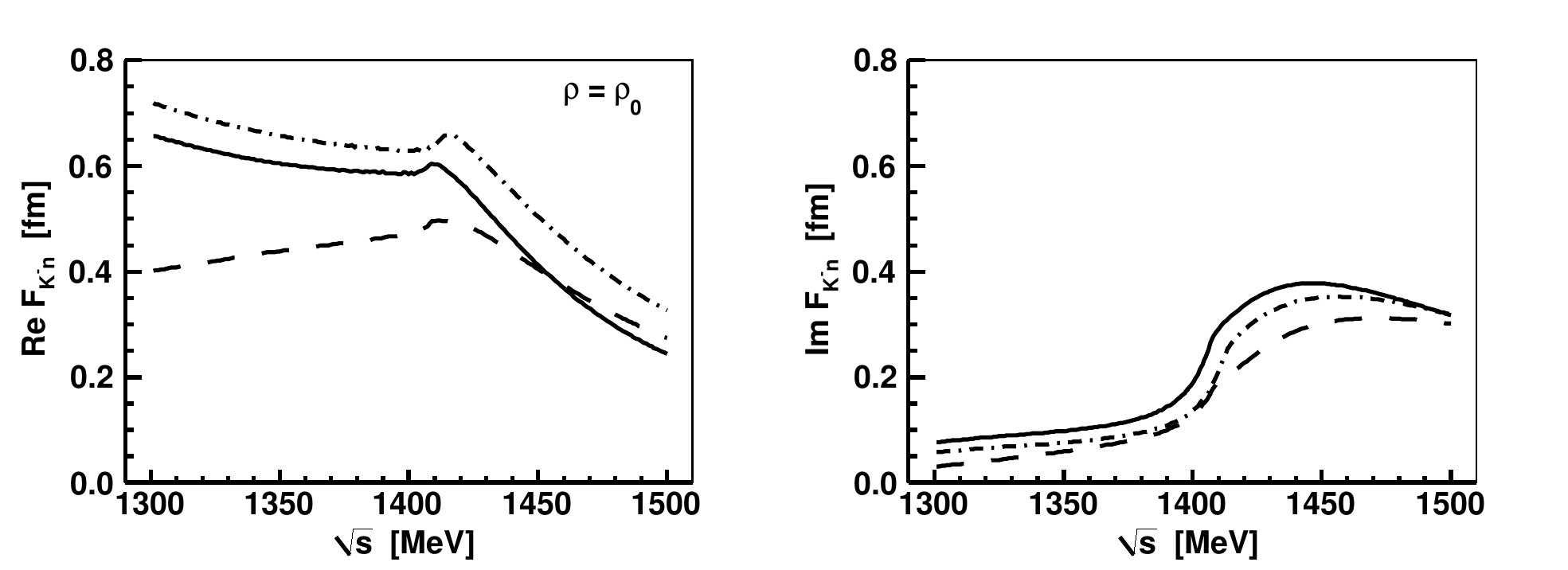}
\end{center}
\caption{Energy dependence of the real (left panels) and imaginary (right panels) parts of 
the elastic $K^-n$ amplitude in the free space (top panels) and in the nuclear medium 
(bottom panels). Dashed curves: TW1 model, dot--dashed curves: CS30 model, 
solid curves: NLO30 model.}
\label{fig:Kn}       
\end{figure}

The energy dependence of the $K^{-}N$ amplitudes in vacuum (nuclear density $\rho = 0$) 
and in nuclear medium ($\rho = \rho_0 = 0.17~{\rm fm}^{-3}$) is shown in figures 
\ref{fig:Kp} and \ref{fig:Kn}. The quantity $\sqrt{s}$ used for the $x$--axis represents 
the meson--baryon energy in the two--body CMS. The nuclear medium affects the $\bar{K}N$ 
interaction in two ways, due to Pauli blocking of the nucleons and due to selfenergies of 
the interacting particles. Both effects are included in our treatment of the 
in--medium amplitudes assuming a symmetric nuclear matter with proton and neutron 
densities $\rho_p = \rho_n = \rho_{0}/2$. It was demonstrated \cite{2011CFGGM} that the 
two effects work partly against each other as the Pauli blocking pushes 
the $\Lambda(1405)$ resonance above the $\bar{K}N$ threshold and the meson 
and baryon selfenergies bring it back to energies about 30 MeV below the threshold. 
As a result the strong energy dependence of the $K^{-}p$ amplitude is preserved 
in nuclear medium too.   

The three models employed in our calculations lead to very similar 
$K^{-}p$ amplitudes above the threshold and are in qualitative agreement at 
subthreshold energies as well. Interestingly, the Figure \ref{fig:Kp} also 
demonstrates that the NLO30 model fitted to the SIDDHARTA data \cite{2011SIDD} 
leads to $K^{-}p$ amplitudes that are in close agreement with those obtained 
for the CS30 model fitted to the DEAR data \cite{2005DEAR}. Especially, the real 
parts of the amplitudes generated by the CS30 and NLO30 models are very close 
to each other when extrapolated as far as to (and even below) the $\pi\Sigma$ 
threshold. This feature may be explained by recalling that in fact the CS30 model 
was not able to reproduce the DEAR data and generates kaonic hydrogen characteristics 
that are more compatible with the SIDDHARTA results. In other words, the energy 
dependence of the $K^{-}p$ amplitude is to some extent fixed by the very precise 
threshold branching ratios and by the low energy scattering and reaction data
(see Ref.~\cite{2006BMN} for a detailed analysis). The significance 
of the new SIDDHARTA data can be judged in terms of putting additional 
constraints on the models and reducing the theoretical uncertainties 
when extrapolating the $K^{-}p$ interaction to subthreshold energies \cite{2012IHW}.
In nuclear medium, the qualitative behavior of the $K^{-}p$ amplitude is once 
again independent of the model used in the calculations. It is remarkable that 
the selfconsistent treatment leads to even smaller differences between the CS30 and NLO30 
models than they were in the free space, especially in the subthreshold energy region.

The $K^{-}n$ elastic scattering amplitudes are shown in Fig.~\ref{fig:Kn}. The cusp 
structure observed for the free--space amplitude is related to the opening of the $K^{-}n$ 
channel and to an existence of a pole in the scattering $S$-matrix, generated dynamically 
at the [$+$,$-$] Riemann sheet (physical in the $\pi\Sigma$ channel and unphysical 
in the $\bar{K}N$ channel). Unlike for the $K^{-}p$ amplitude the CS30 and NLO30 models 
lead to a different behavior of the real part of the $K^{-}n$ amplitude in between 
the $\pi\Sigma$ and $\bar{K}N$ thresholds. This difference is caused partly by different 
positions of the isovector pole (with the pole generated for the CS30 model being closest 
to the physical region and to the $K^{-}n$ threshold) and partly due to larger theoretical 
variations of the computed $K^{-}n$ amplitude since the models are fixed solely 
to the $K^{-}p$ data, not to the $K^{-}n$ ones (since they are not any). In accordance 
with our observations for the $K^{-}p$ system the nuclear medium effects (mainly 
the selfconsistent treatment of the kaon selfenergy) diminish the differences between 
the NLO30 and CS30 models. A shift of the cusp structure to lower energies is also consistent 
with our understanding of the in--medium dynamics. Finally, one should note the different scale 
of the $y$--axis in Figure \ref{fig:Kn} when compared with Figure \ref{fig:Kp}. If the $K^{-}n$ 
amplitude were plotted in the same scale as the $K^{-}p$ one the energy dependence 
of the $K^{-}n$ amplitude would appear much flatter, quite in line with the general consensus 
that the low energy $\bar{K}N$ interaction is governed by the isoscalar $\Lambda(1405)$ 
resonance and the energy dependence of the isovector $\bar{K}N$ amplitude is much weaker.

For a reference we also show in Table \ref{tab:poles} the positions of the poles of the scattering 
$S$--matrix that govern the low energy $K^{-}N$ interactions. The two isoscalar poles found 
on the second Riemann sheet [$-$,$+$] are well known and assigned to the $\Lambda(1405)$ 
resonance. The existence of the isovector pole that relates to the structure observed 
in the $K^{-}n$ amplitude is not so well documented. Though, it was already reported 
in Ref.~\cite{2003JOORM}, at the complex energy $z = (1401 - {\rm i}\:40)$ MeV. 
The Table \ref{tab:poles} also illustrates how much model dependent are the exact positions of the poles.

\begin{table}
\caption{The positions of the poles of the scattering $S$--matrix. The two isoscalar poles $z_1$ and 
$z_2$ are on the [$-$,$+$] Riemann sheet (the signs relate to those of the imaginary parts of the meson--baryon 
CMS momenta in the channels $\pi\Sigma$ and $\bar{K}N$ in this order), the isovector pole $z_3$ 
resides on the [$+$,$-$] Riemann sheet.}
\label{tab:poles}       
\begin{center}
\begin{tabular}{cccc}
\hline\noalign{\smallskip}
model & $z_1$ [MeV] &  $z_2$ [MeV] &  $z_3$ [MeV]  \\
\noalign{\smallskip}\hline\noalign{\smallskip}
TW1   & $(1371 - {\rm i}\:54)$ & $(1433 - {\rm i}\:25)$ & $(1383 - {\rm i}\:53)$ \\
NLO30 & $(1355 - {\rm i}\:86)$ & $(1418 - {\rm i}\:44)$ & $(1410 - {\rm i}\:38)$ \\
CS30  & $(1398 - {\rm i}\:51)$ & $(1441 - {\rm i}\:76)$ & $(1416 - {\rm i}\:24)$ \\
\noalign{\smallskip}\hline
\end{tabular}
\end{center}
\end{table}

The energy and density dependence of the $K^{-}N$ amplitudes plays a key role in construction 
of the $K^{-}$--nuclear optical potential. Considering only one nucleon interactions we have 
$$
V^{K^{-}}_{\rm opt}(\sqrt{s},\rho) \sim F_{K^{-}p}(\sqrt{s},\rho)\: \rho_{p} 
+ F_{K^{-}n}(\sqrt{s},\rho)\: \rho_{n} \;\;\; .
$$
It was demonstrated in Ref.~\cite{2011CFGGM} that the kaon--nucleon CMS energy $\sqrt{s}$ is shifted 
to lower energies with respect to the kaon--nuclear CMS energy due to the binding 
of the hadrons in nuclear matter and due to their motion in the many body system.
Thus, the $K^{-}$--nuclear interaction probes subthreshold $\bar{K}N$ energies where 
the $K^{-}p$ in--medium amplitude exhibits much stronger attraction and the 
resulting $K^{-}$--nuclear optical potential becomes much deeper than 
when it were constructed from the amplitudes taken at the $\bar{K}N$ threshold.
The energy shift to subthreshold energies provides a link between the shallow $\bar{K}$--nuclear 
potentials based on the chiral $\bar{K}N$ amplitudes evaluated at threshold 
and the deep phenomenological optical potentials obtained in fits to kaonic atoms data. 
The relevance of this finding to an analysis of kaonic atoms and quasi--bound 
$\bar{K}$--nuclear states was already investigated in Refs.~\cite{2012FG} and \cite{2012GM}.

We conclude that several versions of coupled--channels separable potential models considered 
in our work provide $\bar{K}N$ amplitudes that exhibit very similar energy dependence in the free 
space as well as in the nuclear medium. Specifically, the strong subthreshold energy 
and density dependence of the $K^{-}p$ amplitudes, that reflects the dominant effect 
of the $\Lambda(1405)$ resonance, does not depend much on a particular version of the model. 
A prominent feature of the models is a sharp increase of $K^{-}p$ in--medium attraction below 
the $\bar{K}N$ threshold that leads to much deeper $\bar{K}$--nuclear optical potentials 
than those that were derived from the amplitudes evaluated at the $\bar{K}N$ threshold.
 
\vspace*{2mm}
\textbf{Acknowledgement:} The work was supported partly by the Grant Agency of Czech Republic,
Grant No.~P203/12/2126.

\end{document}